\documentclass{article}

\usepackage{graphicx}

\usepackage[figuresright]{rotating}

\newcommand{\AmS}{{\protect\the\textfont2

  A\kern-.1667em\lower.5ex\hbox{M}\kern-.125emS}}

\begin{document}

\begin{flushright}

 UND-HEP-06-BIG01

\vspace{2pc}

\end{flushright}

\begin{center}
\bf{Charged Lepton Radiative  and B-meson Double Radiative 
     \\  Decays in Models with Universal Extra Dimensions}

\vspace{2pc}
\end{center}

\begin{center}

Ikaros I. Big$ \mbox{i}^a$, George Chiladz$\mbox{e}^b$, 
Gela Devidz$\mbox{e}^b$, 
Christoph Hanhar$\mbox{t}^c$,\\ 
Akaki Lipartelian$\mbox{i}^b$, Ulf-G. Mei$\ss$ne$\mbox{r}^{c,d}$ \\

\vspace{1pc}

\end{center}

a) University of Notre Dame, Department of Physics, Notre Dame, IN 46556,USA; 
ibigi@nd.edu  \\

b) Institute of High Energy Physics and Informatization, Tbilisi State
University, 
University St.9, Tbilisi 0186, Georgia;\\ chila@hepi.edu.ge; 
devidze@hepi.edu.ge; lipart@hepi.edu.ge \\

c) Institut fur Kernphysik (Theorie), Forschungszentrum Juelich, 52425, 
Juelich, Germany; c.hanhart@fz-juelich.de \\

d) Universitat Bonn, Helmhotz-Institut fur Strahlen- und Kernphysik (Theorie), 
D-53115, Bonn, Germany; meissner@itkp.uni-bonn.de \\

\vspace{2pc}

This paper addresses the role of Large Extra Dimensions in some flavor 
changing neutral current (FCNC) driven processes. In particular we have 
investigated radiative decays of charged leptons within models with only 
one universal extra dimension (UED). Loop contributions with internal 
fermions and scalars of comparable mass would seem to yield sizeable 
amplitudes, since the generic quadratic suppression factor is changed into 
a linear one. Such scenarios can in principle be realized in models with 
universal extra space dimensions. Yet our calculations of the Kaluza-Klein 
(KK) contributions to these radiative decays show this expected relative 
enhancement to disappear due to the near mass degeneracy of the heavy
neutrinos from the different generations. In this paper we  estimate also 
the UED contribution to the $B_{s,d}\rightarrow\gamma\gamma$   rate and 
find an enhancement of 3\% and 6\%, respectively, over the SM prediction.

\newpage

\section{Introduction}

The highly speculative idea of postulating extra dimensions to explain
peculiar features observed in our world with 3+1 space-time dimensions 
has been revived for a novel reason, namely to provide an alternative 
approach to the gauge hierarchy problem [1,2].  An interesting feature  
of this novel insight into the hierarchy problem is that gravitational 
interactions become sizable not at the Planck scale, but already at the 
immensely lower scale $\sim\mbox{O(TeV)}$, which in fact must be considered 
as  the only fundamental scale of nature. TeV scale dynamics in general will 
be explored directly by the LHC starting in 2007. The renaissance of 
multidimensional models is mainly due to superstring theories and their 
generalization, M-theory.  It is the only consistent  quantum theory known 
today that incorporates, at least in principle, all interactions including 
gravity. Both superstring and M-theory most naturally are formulated 
in $d=10$ and $d=11$ dimensions.

Since the extra dimensions can possess very different characteristics, models 
involving them lead to vastly different phenomenologies. Those characteristics 
refer not only to the size and other topological features of the extra
dimensions (and whether they are of the space or time variety), but also to
the fields that populate them.  The options range from where only  the
graviton can propagate through the extra dimenion(s) to where all fields can; 
in the latter case one talks about universal extra dimensions (UED) [3].

A remarkable feature of UED models [3] is the conservation of  the 
so-called Kaluza-Klein (KK) parity, which leads to the absence of tree-level 
KK contributions to transitions at low energies, namely at scales $\mu \ll
1/R$ with $R$ denoting the compactification radius for extra dimensions. KK
parity  resembles R parity, which is conserved in many supersymmetric models. 
In particular KK parity prohibits the production of single KK modes off the 
interaction of ordinary particles.

Transitions driven by FCNC like $K^0-{\bar K}^0$, $B^0-{\bar B}^0$
oscillations and $B_{s,d}\rightarrow\gamma\gamma$ 
are highly suppressed in the 
Standard Model (SM). 
Radiative $\tau$ and $\mu$ decays are  even SM forbidden. New 
Physics contributions in general and KK ones in particular thus have 
(in principle) a considerably enhanced chance to make their presence felt in 
such processes.

In this paper we investigate lepton flavor violating radiative decays of
charged leptons within models with  only one universal extra dimension and  
also estimate their contributions to  $B_{s,d}\rightarrow\gamma\gamma$
transitions, which are allowed though suppressed in the SM. The article is 
organized as follows: after summarizing in Section 2 information about the 
UED model of Appelquist, Cheng and Dobrescu (ACD) [3] relevant for our 
calculations we devote Sections 3 and 4 to the study of charged lepton decays 
$l_i \to l_j\gamma$   and  $B_{s,d}\rightarrow\gamma\gamma$, respectively, 
in the 
framework of the same model, before formulating our conclusions in Section 5. 
Some useful formulas are collected in the Appendix.

\section{The ACD model}

Modern models with extra space-time dimensions have received a great deal of 
attention because the scale at which extra dimensional effects become relevant 
could be around a few TeV [1,2,3]. If so, they could be searched for directly 
at the LHC. The first proposal for using large (TeV) extra dimensions in the
SM with gauge fields in the bulk and matter localized on the orbifold fixed 
points was developed in Ref. [4]. The models  with extra space-time dimensions 
can be built in several ways. Among them the following approaches are the most 
actively pursued ones: i) The ADD model of Arcani-Hammed, Dimopoulos and 
Dvali [1], where all elementary fields except the graviton are localized on a 
brane, while the graviton propagates in the whole bulk. ii) The RS model of 
Randall and Sundrum with warped 5-dimensional space-time and nonfactorized 
geometry [2]. iii) The ACD model of Appelquist, Cheng and Dobrescu (also
referred to as model with Universal Extra Dimensions (UED)), where all fields 
can move in the whole bulk [3].

In UED scenarios the SM fields are thus described as nontrivial functions of
all space-time coordinates. For bosonic fields one simply replaces all
derivatives and fields in the SM Lagrangian by their 5-dimensional
counterparts. These are the $U(1)_Y$  and $SU(2)_L$  gauge fields as well as 
the $SU(3)_C$ gauge fields from the QCD  sector. The Higgs doublet is chosen 
to be even under $P_5$ ($P_5$ is the parity operator in the five dimensional 
space) and possesses a zero mode. Note that all zero modes remain massless
before the Higgs mechanism is applied. In addition we should note that as a 
result of action of the parity operator the fields receive additional masses 
$\sim n/R$ after dimensional reduction and transition to the four dimensional 
Lagrangians; $n$ is a positive integer denoting the KK mode.

In the five dimensional ACD model the same procedure for gauge fixing is
possible as in the models in which fermions are localized on the 4-dimensional 
subspace. With the gauge fixed, one can diagonalize the kinetic terms of the 
bosons and finally derive expressions for the propagators. Compared to the SM, 
there are additional Kaluza-Klein (KK) mass terms. As they are common to all 
fields, their contributions to the gauge boson mass matrix is proportional to 
the unity matrix. As a consequence, the electroweak angle remains the same for 
all KK-modes and is the usual Weinberg angle  $\theta_W$. Because of the 
KK-contribution to the mass matrix, charged and neutral Higgs components with 
$n\not=0$  no longer play the role of Goldstone bosons. Instead, they mix with 
$W_5^{\pm}$  and $Z_5$ to form - in addition to the Goldstone modes $G^0_{(n)}$  and 
$G^{\pm}_{(n)}$ - three physical states $a^0_{(n)}$  and $a^{\pm}_{(n)}$. 
Below we will study the impact of these new {\it charged} states.

The interactions of ordinary charged leptons with  pairs of KK scalars and 
neutrinos ($a_{(n)}$, $\nu_{k(n)}$) is given by 

\begin{equation}
L(l_j \nu_{i(n)} a^+_{(n)})=\bar{l_j}(c_L P_L + c_R P_R)
U_{ij}\nu_{i(n)}a^+_{(n)}+h.c.            
\label{LAG}
\end{equation}
with 
$$
P_{R,L}=\frac{1\pm\gamma_5}{2},\hspace{6mm} 
c_L=-(g_2nm(l_j))/(\sqrt2 M_{W(n)}), \hspace{6mm}
c_R=-(g_2 M_W)/(\sqrt2 M_{W(n)})\\
$$

where $g_2$ is the $SU(2)$ coupling for weak interaction, $n$ labels 
the KK towers  (e.g.  $M_{W(n)}$ is the mass for  
$n$-th KK-mode:  $M^2_{W(n)}=M^2_W+n^2/R^2$);  
$U_{ij}$ is an element in the MNS matrix, 
the leptonic analogue of the CKM   
matrix.

The complete list of Feynman rules for models with only one universal 
extra dimension has been given in Ref. [5].

The Lagrangian responsible for the interaction of the charged scalar KK 
towers $a^{*}_{(n)} $  with the ordinary down quarks, is as follows 

\begin{equation}
{\mathcal L}=\frac{g_2}{\sqrt2 M_{W(n)}}\bar{Q}_{i(n)}
(C_L^{(1)}P_L+C_R^{(1)}P_R)a^*_{(n)}d_j+
  \frac{g_2}{\sqrt2M_{W(n)}}\bar{U}_{i(n)}
(C_L^{(2)}P_L+C_R^{(2)}P_R)a^*_{(n)}d_j~,        
\end{equation}
%
In the equation (2) the following notations are used [5]: 
\begin{eqnarray}
C_L^{(1)}&=&-m_3^{(i)}V_{ij}, \hspace{9mm}
C_L^{(2)}=m_4^{(i)}V_{ij},  \nonumber\\
C_R^{(1)}&=&M_3^{(i,j)}V_{ij}, \hspace{2mm
C_R^{(2)}=-M_4^{(i,j)}V_{ij}}, \nonumber\\
M^2_{W(n)}&=&m^2(a^*_{(n)})=M^2_W+\frac{n^2}{R^2},        
\end{eqnarray}
where $V_{ij}$ are elements of the CKM matrix. The mass parameters 
in Eq.(3) are defined as
\begin{eqnarray}
m_3^{(i)}&=&-M_Wc_{i(n)}+\frac{n}{R}\frac{m_i}{M_W}s_{i(n)},\hspace{5mm}
m_4^{(i)}=M_Ws_{i(n)}+\frac{n}{R}\frac{m_i}{M_W}c_{i(n)},\nonumber\\
M_3^{(i,j)}&=&\frac{n}{R}\frac{m_j}{M_W}c_{i(n)},\hspace{35mm}
M_4^{(i,j)}=\frac{n}{R}\frac{m_j}{M_W}s_{i(n)}.     
\end{eqnarray}

Here, $M_W$ and the masses of up (down)-quarks $m_i \, (m_j)$ on the 
Right hand side of Eq.(4) are zero mode masses and the $c_{i(n)}$, $s_{i(n)}$ 
denote the cosine and sine of the fermion mixing angles, respectively
\begin{equation}
\tan 2\alpha_{f(n)}=\frac{m_f}{n/R},\hspace{3mm} n\geq1~.   
\end{equation}
The masses for the fermions are calculated as
\begin{equation} 
m_{f(n)}=\sqrt{\frac{n^2}{R^2}+m^2_f} . 
\end{equation}

In the phenomenological applications we use the restriction 
$n/R \geq 250\,\mbox{GeV}$ and hence we assume that all the fermionic
mixing angles except $\alpha_{t(n)}$ are equal zero.


\section{Radiative decays $l_i\to l_j \gamma$ of Charged Leptons}

In Ref.[6] a mainly model independent analysis of $\mu\to e\gamma$  and  
$\tau\to \mu\gamma$ on the one-loop level has been given. An important 
statement is that when the masses of the internal scalar and fermion  
masses are comparable, the transition amplitude becomes considerably 
enhanced over what one would usually expect. The authors state that 
such a situation arises in UED theories. We will analyze the situation 
in more detail specifically in theories with only one extra dimension [3].

Real decay processes, where the photon has to be on-shell represent a 
magnetic transition described by two form factors:

$$
A(l_i\rightarrow l_j\gamma)=\epsilon^{\mu}(k)\bar{u_j}(p_2)
i\sigma_{\mu\nu}k^{\nu}(F_{2V}+ 
F_{2A}\gamma_5)u_i(p_1)    \; ;     \eqno{(7)}
\label{REALAMP}                              
$$
$u_{i,j}$ denote the lepton spinors with momenta $p_i$ and $k = p_1 - p_2$ and 
$\epsilon^{\mu}(k)$ the photon polarization vector.

For completeness we give also the amplitude for such transitions with an 
{\cal off} shell photon (or $Z^0$), which contains four additional form 
factors: 
$$
A(l_i\rightarrow l_j\gamma)=\epsilon^{\mu}(k)\bar{u_j}(p_2)
[(F_{1V}+F_{1A}\gamma_5)\gamma_{\mu}+ 
$$
$$
i\sigma_{\mu\nu}k^{\nu}(F_{2V}+F_{2A}\gamma_5)+
k_{\mu}(F_{3V}+F_{3A}\gamma_5)]
u_i(p_1) \eqno{(8)}
\label{FULLAMP}                         
$$

The specifics of the underlying dynamics then determine the form factors 
$F_{iV}$ and $F_{iA}$. 
`Switching on' one universal extra dimension expands the particle
 content of the model. In particular, KK-modes of the charged scalar 
boson  are real particles in this case[3,5]. 
The relevant Feynman diagrams are shown in 
Fig.1.

\begin{figure}
\centering \includegraphics[width=60mm]{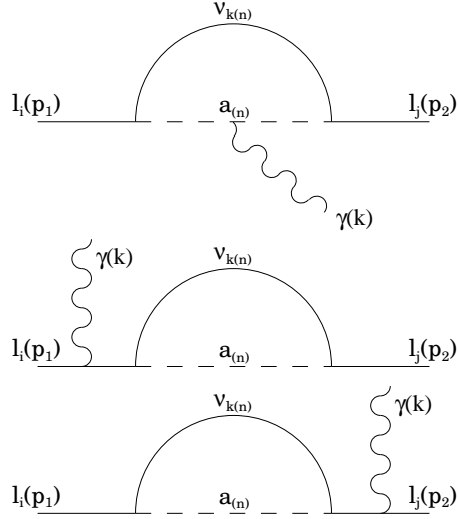}
\caption{$l_i\rightarrow l_j\gamma$ decay in the model with Universal
  Extra Dimension}
\label{fig:toosmall}
\end{figure}

Their contributions to $l_i\rightarrow l_j\gamma$ are explicitly calculated 
in this section.

One can conclude from the expressions for the masses of the neutrino and 
physical scalar `towers'
$$
m^2(\nu_{k(n)})=m^2(\nu_{k})+\frac{n^2}{R^2}, \hspace{10mm}
m^2(a_n)=M^2_W+\frac{n^2}{R^2}   \eqno{(9)}      
\label{TOWERS}
$$
that they are comparable already for the first excited
mode ($ n=1$), because very rough estimates for the compactification 
radius  tell us: $1/R>250 GeV$.\\
Explicit calculations yield for the form factors $F_{2V,2A}$ - relevant 
for $l_i \to l_j \gamma$ - and the other four form factors $F_{1V,1A}$ and 
$F_{3V,3A}$ (which come relevant for $l_i\to l_jl\bar l$)

$$
F_{1V}=-
\frac{i}{(4\pi)^2}eg^2_2 \frac{M^2_W}{M^2_{W(n)}}
\frac{k^2}{m^2(a_{(n)})}U_{ik}U^*_{jk}
\biggl{\{} \frac{nm_{\nu_{k(n)}}}{RM_W^2} 
\frac{(m_{l_i}+m_{l_j})^2}{2m^2(a_{(n)})}f_1(x_k)+ \hspace{3mm}(10\mbox{a})
$$
$$
\biggl(1+\frac{n^2}{R^2M_W^2}\frac{m_{l_i}m_{l_j}}{M^2_W}\biggr)
\biggl( \frac{1}{4}f_2(x_k)+  
\frac{m^2_{l_i}+m^2_{l_j}+m_{l_i}m_{l_j}}{m^2(a_{(n)})}f_3(x_k)+
\frac{1}{6}\frac{k^2}{m^2(a_{(n)})}
f_4(x_k)\biggr)
 \biggr{\}}
 \label{F1FF}
$$
$$
F_{2V}=-\frac{1}{2}
\frac{i}{(4\pi)^2}eg^2_2 \frac{M^2_W}{M^2_{W(n)}}
\frac{m_{l_i}+m_{l_j}}{m^2(a_{(n)})}U_{ik}U^*_{jk}
\biggl{\{}  \frac{n}{RM_W} \frac{m_{\nu_{k(n)}}}{M_W}\biggl(
-f_5(x_k)+ 
$$
$$
\frac{k^2}{m^2(a_{(n)})}f_1(x_k)-
\frac{m^2_{l_i}+m^2_{l_j}}{m^2(a_{(n)})}f_6(x_k)\biggr)+
\biggl(1+\frac{n^2}{R^2M_W^2}\frac{m_{l_i}m_{l_j}}{M^2_W}\biggr)
\biggl( -\frac{1}{2}f_7(x_k)+
$$
$$
\frac{k^2}{m^2(a_{(n)})} f_3(x_k)-
\frac{2m^2_{l_i}+2m^2_{l_j}+3m_{l_i}m_{l_j}}{3m^2(a_{(n)})}f_4(x_k)
 \biggr) \biggr{\}}\hspace{10mm} (10\mbox{b})
 \label{F2FF}
$$
$$
F_{3V}=-
\frac{i}{(4\pi)^2}eg^2_2 \frac{M^2_W}{M^2_{W(n)}}
\frac{m^2(l_i)-m^2(l_j)}{m^2(a_{(n)})}U_{ik}U^*_{jk}
\biggl{\{}\frac{n}{RM_W} \frac{m_{\nu_{k(n)}}}{M_W}
\frac{(m_{l_i}+m_{l_j})^2}{2m^2(a_{(n)})}f_1(x_k)+
$$
$$
\biggl(1+\frac{n^2}{R^2M_W^2}\frac{m_{l_i}m_{l_j}}{M^2_W}\biggr)
\biggl( \frac{1}{4}f_2(x_k)+
\frac{m^2_{l_i}+m^2_{l_j}+m_{l_i}m_{l_j}}{m^2(a_{(n)})}f_3(x_k)+
\frac{1}{6}\frac{k^2}{m^2(a_{(n)})}
f_4(x_k)
 \biggr) \biggr{\}}
\eqno{(10c)}
 \label{F3FF} 
$$
where 
$$
x_k=m^2(\nu_{k(n)})/m^2(a_{(n)}) \eqno{(11)}
$$
and summation over the  
tower indices $n$ is assumed in Eqs.(10a-10c); 
the functions $f(x_k)$  are given in the appendix. 
The axial form-factors are related with the corresponding 
vector  ones  by: 
$$
F_{A}(m(l_i),m(l_j))=F_{V}(m(l_i),-m(l_j)) \eqno{(12)}
$$

Eqs.  (10a-10c) demonstrate explicitly the general relation 
between form factors:
$$
(m(l_i)-m(l_j))F_{1V}=-k^2F_{3V}  \eqno{(13)}
$$
Using Eq.({\ref{REALAMP}) we obtain for the decay width:
$$
\Gamma(l_i\rightarrow l_j\gamma )=
\frac{\mid F_{2V}\mid^2+\mid F_{2A}\mid^2} 
{8\pi} \Biggl( \frac{m^2_{l_i}-m^2_{l_j}}{m_{l_i}}\Biggr)^3 \eqno{(14)}
\label{WIDTH}
$$
Let us note that the ratio $x_k=m^2(\nu_{k(n)})/m^2(a_{(n)})$ is close 
to unity for all $n$, namely $0.9<x_n<1$): 
$$
x_k=\frac{m^2(\nu_{k(n)})}{m^2(a_{(n)})}=
\frac{m^2(\nu_k)+\frac{n^2}{R^2}}
{M^2_W+\frac{n^2}{R^2}}  \eqno{(15)}
$$
With the rough estimate for the compactification
 radius ($1/R>250GeV$) we have already for the first 
KK-mode  $x_k>0.9$. Noting that the masses of the scalar and fermion 
towers are 
close to each other for the same $n$, we can simplify Eq.(10b):
$$
F_{2V}=\frac{1}{2}
\frac{i}{(4\pi)^2}eg^2_2 \frac{M^2_W}{M^2_{W(n)}}
\frac{m(l_i)}{m^2(a_{(n)})}U_{ik}U^*_{jk}
\biggl(  \frac{n}{RM_W} \frac{m_{\nu_{k(n)}}}{M_W}\bigl(
-\frac14+  \frac{x_k}{12}\bigr)+
\frac{x_k}{60}\biggr) \eqno{(16)}
\label{F2FFSIMP}
$$
and thus 
$$
Br(l_i\rightarrow l_j\gamma)=\frac{6\alpha}{\pi}
\frac{M^8_W}{M^8_{W(n)}}\bigl(U_{ik}U^*_{jk}f(x_k)\bigr)^2 \eqno{(17)}
\label{BR}
$$
In four-dimensional models with small Dirac neutrino masses the ratio 
of neutrino mass square differences to the $W$-boson square mass 
serves as a highly efficient {\em suppression} factor for 
$l_i\rightarrow l_j\gamma$.  
Eq. (\ref{F2FFSIMP}) exhibits an apparently {\em linear} dependence on 
the neutrino 
to $W$ mass ratio for the exchange of KK towers in the loops. 
This might be seen at first as leading to a very considerable enhancement 
of the 
$l_i\rightarrow l_j\gamma$  amplitude. This conclusion, however, would be 
fallacious. 
For upon explicit substitution of Eq.(\ref{TOWERS}) for the KK masses 
into the form factor expression in Eq.(\ref{F2FFSIMP}) the {\em quadratic} 
dependence re-emerges: 
$$
\frac{n}{RM_W}U_{ik}U^*_{jk}\frac{m(\nu_{k(n)})}{M_W}=
\frac{n}{RM_W}U_{ik}U^*_{jk}\bigl( \frac{n}{R}+
\frac{Rm^2(\nu_k)}{2n}\bigr) \frac{1}{M_W}=
U_{ik}U^*_{jk}\frac{m^2(\nu_k)}{2M^2_W} \; . \eqno{(18)} 
\label{QUAD} 
$$

The initial appearance of a merely linear suppression thus disappears due to 
the near-degeneracy of the masses for 
neutrino KK-towers from different generations. 
For example, we have for two neutrino generations :

$$      
m(\nu_{\mu(n)})-m(\nu_{e(n)})=
(m(\nu_{\mu})-m(\nu_{e}))
\frac{m(\nu_{\mu})+m(\nu_{e})}{2n/R}
\ll 
m(\nu_{\mu})-m(\nu_{e}) \eqno{(19)}
$$

In the end the following expression emerges for the branching ratio:

$$
Br(l_i\rightarrow l_j\gamma)=\frac{3\alpha}{32\pi}
\frac{M^8_W}{M^8_{W(n)}}
\biggl(U_{ik}U^*_{jk}\frac{m^2(\nu_k)}{M^2_W}\biggr)^2 \eqno{(20)}
\label{BRLED} 
$$

This expression shows that it cannot enhance
 the SM result [7-12]:                       

$$
Br(l_i\rightarrow l_j\gamma)_{SM}=\frac{3\alpha}{32\pi}
\biggl(U_{ik}U^*_{jk}\frac{m^2(\nu_k)}{M^2_W}\biggr)^2 \eqno{(21)}
\label{BRSM}
$$

In the slightly extended SM with massive left-handed neutrinos lepton 
flavor violating processes like $\mu\rightarrow e\gamma$  and 
$\tau\rightarrow \mu\gamma$ are extremelly  suppressed. 
For example, taking into consideration the data from modern neutrino 
experiments [13], the branching ratio for  
$\mu\rightarrow e\gamma$ is predicted to be as low as $10^{-57}$ [7-12] 
in the SM.      
Planned experiment are expected to lower the existing upper bound 
$Br(\mu\rightarrow e\gamma)_{exp}<1.2\cdot 10^{-11}$
to the $10^{-13}\div 10^{-14}$ levels [14]. 
\vspace{1pc}


\section{ $B_{s,d}\rightarrow\gamma\gamma$  decay in models with one 
universal extra dimension}

Detailed studies of the decays of strange hadrons were instrumental in 
developing the Standard Model (SM). Recent findings on B decays - in 
particular the CP asymmetry observed in 
$B\to \psi K_{S}$ by the BABAR and BELLE collaborations at the $B$ factories - 
represent a striking confirmation of the SM[15]. 
Yet they do not invalidate the various theoretical arguments pointing to 
its incompleteness, i.e. the existence of  physics beyond the SM (BSMP). 
If anything they even strengthen the case for a new paradigm. History might 
well repeat itself in the sense that future detailed studies of the decays 
of beauty and charm hadrons and tau leptons will reveal the intervention 
of BSMP. 

The BABAR and BELLE experiments running at the two B factories at SLAC in 
the USA and at KEK in Japan are producing the huge high-quality data sets 
required for such searches for BSMP. There is even a proposal in Japan for 
building a Super-B factory with much higher luminosity; similar plans are 
being pursued in Italy. Further information will be gained from the tau-charm 
factories at Cornell University in the US and at Beijing in China. 

Some experimental evidence for an incompleteness of the CKM description has 
actually emerged in the CP asymmetry in  $B\to \psi K_{S}$   and similar 
channels. It also points towards radiative and related B decays as promising 
areas for search for BSMP. 

       Exploration of B-physics, including B-meson rare decays is one of the 
central issues of the physical programs at accelerator  facilities operating 
now or soon getting online. The process $B_{s,d}\rightarrow\gamma\gamma$  , 
which is the subject of this section,  has an  unusual experimental signature 
that can be searched for at least at B factories. It should be noted that the 
two photon final states produced in this process can be CP even as well as CP 
odd. This feature might allow searches for nontraditional sources of CP 
violation in B-physics. In general  the process 
$B_{s,d}\rightarrow\gamma\gamma$  is sensitive to BSMP effects. 
The experimental feasibility has stimulated efforts of theoretical groups as  
well [16-33]. $B_{s,d}\rightarrow\gamma\gamma$  rates have been calculated 
within the SM with and without  QCD corrections, in multi-Higgs doublet as 
well as supersymmetric models.

        In the SM  $B_{s,d}\rightarrow\gamma\gamma$  first arise at the one 
loop level with the exchange in the loops by up-type quarks and
 W-bosons [16-20] . SM predictions for the branching ratios are of the order 
of  $10^{-7}(10^{-9})$.

       It has been shown that in extended versions of the SM (multi-Higgs 
doublet and  supersymmetric models) one could reach branching  ratios as 
large as   $Br(B_{s,d}\rightarrow\gamma\gamma)\sim 10^{-6}$  
depending on the parameters of the models. This enhancement was achieved 
mainly due to exchange of charged scalar Higgs particles within the loop. 
There exists an analogous possibility in other exotic models as well for the 
scalar particle exchange inside the loop, which could enhance this process. 
For example, the ACD model with only one 
universal extra dimension [3] presents us with such an  opportunity. One 
should note that in the above approach towers of charged Higgs particles 
arise as real objects with definite masses, not as fictitious (ghost) fields.

In this Section we calculate the contributions from these real scalars to 
$B_{s,d}\rightarrow\gamma\gamma$.  
The Feynman graphs, describing the contributions of scalar physical
states to process under consideration, are shown in Fig.2. 

\begin{figure}[htb]
\centering 
\includegraphics[width=60mm]{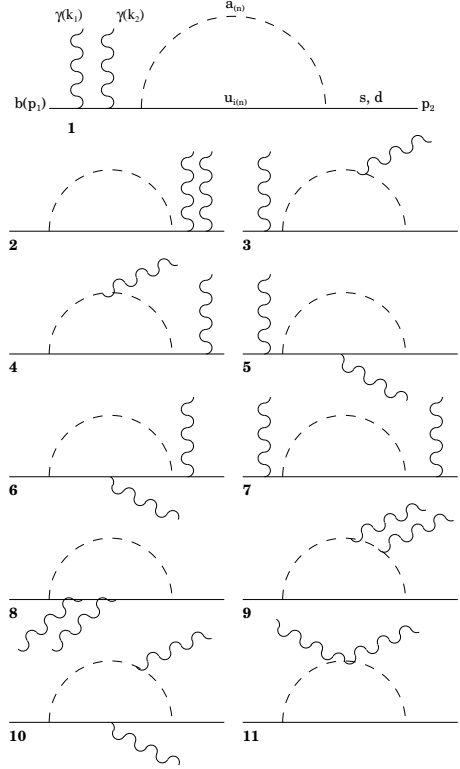}
\caption{Double radiative $B$-meson decay
  $B_{s,d}\rightarrow\gamma\gamma$ 
in the theory with only one extra universal dimension (the dashed lines in
  the loops correspond to the charged KK towers $a^*_{(n)}$, while the
  solid lines in the loops are for the up-quark KK towers).}
\label{fig:toosmall}
\end{figure}

The amplitude for the decay 
$B_{s,d}\rightarrow\gamma\gamma$ has the form
$$
T(B\rightarrow\gamma\gamma) =\epsilon_1^{\mu}(k_1)\epsilon_2^{\nu}(k_2) 
[Ag_{\mu\nu} + 
iB\epsilon_{\mu\nu\alpha\beta}k_1^{\alpha}k_2^{\beta}]. \eqno{(22)}
$$

This equation holds after gauge fixing for the final photons 
which we have chosen as

$$
\epsilon_1 \cdot k_1=\epsilon_2 \cdot k_2=\epsilon_1 \cdot k_2
=\epsilon_2 \cdot k_2=0, \eqno{(23)}
$$
where $\epsilon_1$ and $\epsilon_2$ are photon polarization vectors, 
respectively.
The gauge condition Eq.(23) together
with  energy-momentum conservation leads to

$$
\epsilon_i \cdot P=\epsilon_i \cdot p_1=\epsilon_i\cdot p_2=0, \eqno{(24)}
$$
where

$$
P=k_1+k_2 \hspace{3mm} \mbox{and}\hspace{3mm} 
p_1=p_2+k_1+k_2.  \eqno{(25)}   
$$                    


Let us write down some useful kinematical relations resulting
 from Eqs.(23-25):

$$
P\cdot p_1 =m_bM_B,\quad  P\cdot p_2 =-m_{s(d)}M_B, \quad 
P\cdot k_1 = P\cdot k_2 = k_1\cdot k_2 =\frac12M^2_B, 
$$
$$
p_1 \cdot p_2 =-m_bm_{s(d)},\quad
p_1\cdot k_1 = p_1 \cdot k_2 = \frac12m_bM_B,
$$
$$
p_2 \cdot k_1 = p_2 \cdot k_2 =-\frac12m_{s(d)}M_B. \eqno{(26)}
$$

The total contributions into $CP$-even ($A$) and $CP$-odd ($B$)
 amplitudes from 
Eq.(22) are calculated as sums of the  appropriate contributions
of the diagrams in Fig.2 corresponding to a tower of scalar particle
contributions in the ACD model with only one extra dimension.
Let us note that we used the following formula for the hadronic 
matrix elements:
$$
\langle 0 | {\bar s\,(\bar d)} \gamma_{\mu}\gamma_5 b | B(P)\rangle=
-if_BP_{\mu}.  \eqno{(27)} 
$$
Apart from one particle reducible (1PR) diagrams, one particle
irreducible (1PI) ones contribute to the amplitudes, and hence, 
to their $CP$-even ($A$) and  $CP$-odd ($B$) parts. 
We should note that each of 
the 1PI contributions is finite. Let us discuss these contributions 
in more details. In the SM only one 1PI diagram (the one with the $W$-boson 
exchange in the loop, when both photons are emitted by virtual
up-quarks) gives a contribution of the order of
$\sim 1/M^2_W$. In the Ref.[34]
it was observed that diagrams with  light quark exchanges contribute 
as $\sim1/M^2_W$, 
while diagrams containing the heavy quarks are of order  $\sim1/M^4_W$. 
In the ACD  model the contributions of such diagrams are also of  order
$\sim1/M^4_W$ because 
the estimate for all KK-tower masses, including the ones exchanged in 
the loops, is given in units $1/R\geq 250\,$GeV [35,36]. 
Similar considerations show that all
the 1PI diagrams existing in the ACD model also are of order $\sim1/M^4_W$. 
Thus, the leading 1PI diagrams are negligible and we do not consider them.

The total contributions to the $B\rightarrow\gamma\gamma$  decay 
amplitudes are:
$$
A=
\frac14\frac{i}{(4\pi)^2}e^2g^2_2f_B\frac{Q_d}{M^2_{W(n)}}
\frac{V^*_{is(d)}V_{ib}}{m^2(a^*_{(n)})}
\frac{m_b^3}{m_{s(d)}}\Bigl{\{}
\frac{n}{RM_W}m^{(i)}_3m_{i(n)}c_{i(n)}f_8(x_i)+ 
$$
$$
 \bigl[ (m_3^{(i)})^2
-\frac{n^2}{R^2M^2_W}m_bm_{s(d)}c^2_{i(n)}\bigr]\frac12 f_7(x_i)
\Bigr{\}}, 
$$
$$
B=
\frac12\frac{i}{(4\pi)^2}e^2g^2_2f_B\frac{Q_d}{M^2_{W(n)}}
\frac{V^*_{is(d)}V_{ib}}{m^2(a^*_{(n)})}
\frac{m_b}{m_{s(d)}}\Bigl{\{}
\frac{n}{RM_W}m^{(i)}_3m_{i(n)}c_{i(n)}f_8(x_i)+ 
$$
$$
\bigl[ (m_3^{(i)})^2
+\frac{n^2}{R^2M^2_W}m_bm_{s(d)}c^2_{i(n)}\bigr]\frac12 f_7(x_i)
\Bigr{\}}, \eqno{(28)}
$$                        
where
$$
f_8(x)=\frac{-5x^2+8x-3+2(3x-2)\ln x}{6(1-x)^3}, 
$$
$$
f_7(x)=\frac{-2x^3-3x^2+6x-1+6x^2\ln x}{6(1-x)^4},
$$
$$
x_i=\frac{m^2(u_{i(n)})}{m^2(a^*_{(n)})}~.\eqno{(29)}
$$

As  is obvious from Fig.2, the correct calculation must include 
the crossed diagrams (not shown on fig.2).
In the kinematics we use, cf. Eqs.(23-26), this leads to  a factor 2
for all amplitudes, except for the one given by diagram 11.
 However, diagram 11 belongs to the class of 1PI diagrams. As it
was stated above, those contributions are order $\sim1/M^4_W$ and thus
negligible compared to those from the 1PR diagrams.

        On the other hand, using the unitarity of the 
CKM matrix, the amplitude for double radiative
$B$-meson decay can be rewritten as:
$$
T=\sum_{i=u,c,t}\lambda_iT_i=\lambda_t\Bigl\{T_t-T_c+
\frac{\lambda_u}{\lambda_t}(T_u-T_c)\Bigr\}. \eqno{(30)} 
$$

Let us note that we restricted ourselves to calculating the leading 
order terms of $\sim1/M^2_W$ from the up-quark KK-towers. 
In this approximation it 
turns out that the $u_{(n)}$  and the $c_{(n)}$ towers have equal
contributions. 
Therefore, the expressions for the amplitudes have a simpler form than 
before:

$$
A=\lambda_t(A_{t(n)}-A_{c(n)})~,
$$
$$
B=\lambda_t(B_{t(n)}-B_{c(n)})~. \eqno{(31)}   
$$
Furthermore, it is easy to obtain from Eq.(22) the expression 
for the $B\rightarrow\gamma\gamma$ decay partial width:
$$
\Gamma(B\rightarrow\gamma\gamma)=\frac{1}{32\pi M_B}
\Bigl[ 4 | A |^2+\frac12M^4_B |B|^2   \Bigr]. \eqno{(32)}  
$$

Now we are in the position to compare the ACD contribution to the 
decay with that of the SM. 
Namely, let us consider the ratio:

$$
\frac{\Gamma(B_{s(d)}\rightarrow\gamma\gamma)_{\rm ACD}}
{\Gamma(B_{s(d)}\rightarrow\gamma\gamma)_{\rm SM}}=
$$
$$
\frac{24n^2M^6_W}{Q^2_dR^2M^4_{W(n)}m^4(a^*_{(n)})} 
\cdot \frac {
 \Bigl{\{}\frac{m^{(i)}_3m_{i(n)}}{M^2_W}c_{t(n)}f_8(x_{t(n)})
+ \frac{n}{RM_W}f_8(x_{c(n)})\Bigr{\}}^2 
 }
{4\bigl( C(x_t)+\frac{23}{3} \bigr)^2 
+ 2\bigl(C(x_t)+\frac{23}{3} \nonumber\\
+
16\frac{m_{s(d)}}{m_b}\bigr)^2 }
$$
where
$$
C(x)=\frac{22x^3-153x^2+159x-46}{6(1-x)^3} 
+\frac{3(2-3x)}{(1-x)^4}\ln x, 
\hspace{6mm}\qquad x_t=\frac{m^2_t}{M^2_W}.\eqno{(34)} 
$$

Rough numerical estimates show that pure UED contributions to 
$B_{s}\rightarrow\gamma\gamma$ and $B_{d}\rightarrow\gamma\gamma$
enhance the SM estimate by about $\sim 3\%$ and $\sim 6\%$, respectively.

\vspace{1pc}

{\bf Conclusion}\\
\vspace{0.5pc}

In this paper we have investigated lepton flavor violating radiative decays 
of charged leptons  $l_i\to l_j \gamma$
within models with only one universal extra dimension and 
have estimated also their 
contributions to the highly suppressed SM rates for 
$B_{s,d}\rightarrow\gamma\gamma$.
Planned experiment are expected to lower the existing upper bound 
$Br(\mu\rightarrow e\gamma)_{exp}<1.2\cdot 10^{-11}$
 to the $10^{-13}\div 10^{-14}$ levels[14]. There are bad news and good news 
from our analysis: 

\begin{itemize}
\item 
The bad news are that UED models with only one additional spatial dimension 
cannot raise $Br(\mu\rightarrow e\gamma)$ into a range, where it could ever 
be observed. 

\item 
The good news are that if $\mu\rightarrow e\gamma$ is ever observed, it must 
have a completely different origin. 

\end{itemize}

Th pure UED contribution to the $B_{s}\rightarrow\gamma\gamma$
[$B_{d}\rightarrow\gamma\gamma$] rate is $3\%$[$6\%$] 
of the SM estimates of 
$Br(B_{s}\rightarrow\gamma\gamma)\sim 10^{-7}$ and
$Br(B_{d}\rightarrow\gamma\gamma)\sim 10^{-9}$, i.e. rather small. It is quite 
possible that the as yet uncalculated radiative QCD corrections could enhance 
these rates further and that they become observable at a Super-B factory. Then 
they might be relevant for the central goal of B physics studies to not only 
discover New Physics, but also identify its salient features.


\newpage

{\bf Acknowledgements}

\vspace{0.5pc}

 This research was made possible in part
 by Award No. GEP1-33-25-TB-02 of the Georgian Research and
 Development Foundation (GRDF) and the U.S. Civilian
 Research \& Development Foundation for the Independent States of the
 Former Soviet Union (CRDF). Part of this work have been done due to
 support of Deutscher Akademischer Austauchdienst (DAAD).
This work was also supported by the NSF under grant number PHY-0355098, 
as well as this research is part of the EU 
Integrated Infrastructure Initiative Hadron Physics Project
under contract number RII3-CT-2004-506078.
  G.D. and A.L. wish to express their great
 attitude to A.~Kacharava, 
N.N.~Nikolaev,  H.~Str\"oher, 
 A.~Wirzba for valuable discussions and help during our 
 stay at IKP-FZJ as well as for excellent scientific environment and warm
 hospitality they provided for us. 
I.~Antoniadis, A.~Khelashvili, C.~Kolda, T.~Kopaleishvili and G.G.~Volkov are
warmly thanked for sharing their  insight into the topics discussed here.
We are also grateful to D.~Chiladze, 
A.~Garishvili and I.~Keshelashvili for their help and friendly support.

\bigskip

{\bf Appendix}

\vspace{0.5pc}

The functions $f(x_i)$ appearing in Eqs.(10a-10c, 17) 
are given by:  
$$
f_1(x)=\frac{1-9x-9x^2+17x^3-6x^2(3+x)lnx}{36(1-x)^5}
$$
$$
f_2(x)=\frac{2-9x+18x^2-11x^3+6x^3lnx}{18(1-x)^4}\\
$$
$$
f_3(x)=\frac{1-8x+36x^2+8x^3-37x^4+
12x^3(4+x)lnx}{144(1-x)^6}\\
$$
$$
f_4(x)=\frac{3-20x+60x^2-120x^3+65x^4+
12x^5-60x^4lnx}{240(1-x)^6}\\
$$
$$
f_5(x)=\frac{x^2-1-2xlnx}{2(1-x)^3}\\
$$
$$
f_6(x)=\frac{-1-9x+9x^2+x^3-6x^(1+x)lnx}{6(1-x)^5}\\
$$
$$
f_7(x)=\frac{-1+6x-3x^2-2x^3+6x^2lnx}{6(1-x)^4}\\
$$
$$
f(x_k)=\frac{n}{rM_W}\frac{m(\nu_{k(n)})}{M_W}
\Bigl( -\frac14+\frac{x_k}{12}\Bigr)+\frac{x_k}{60} \\
$$


\end{document}